\documentclass{emulateapj}
\usepackage{apjfonts}

\slugcomment{Accepted for the publication in ApJ}
\shorttitle{Stability of MRI-Turbulent Accretion Disks}
\shortauthors{Takahashi \& Masada}
\begin{document}
\title{Stability of MRI turbulent Accretion Disks}
\author{Hiroyuki R. Takahashi\altaffilmark{1}  and Youhei
Masada\altaffilmark{1,2}}
\altaffiltext{1}{National Astronomical Observatory of Japan, Osawa,
Mitaka, Tokyo 181-8588, Japan; takahashi@cfca.jp}
\altaffiltext{2}{Department of Computational Science, Kobe University,
1-1 Rokkodai, Nada, Kobe 657-8501}

\begin{abstract}
Based on the characteristics of the magnetorotational instability (MRI) and
the MRI-driven turbulence, we construct a steady model for a geometrically
thin disk using "non-standard" $\alpha$-prescription. 
The efficiency of the angular momentum transport depends on the magnetic
Prandtl number, $Pm = \nu/\eta$, where $\nu$ and $\eta$ are the
microscopic viscous and magnetic diffusivities. In our disk model,
Shakura-Sunyaev's $\alpha$-parameter has a power-law dependence on
the magnetic Prandtl number, that is $\alpha \propto Pm^\delta$ where
$\delta$ is the constant power-law index.
Adopting Spitzer's microscopic diffusivities, the magnetic Prandtl
number becomes a decreasing function of the disk radius when $\delta > 0$.
The transport efficiency of the angular momentum and the viscous heating
rate are thus smaller in the outer part of the disk, while these are
impacted by the size of index $\delta$. We find that the disk becomes
more unstable to the gravitational instability for a larger value of
index $\delta$. 
The most remarkable feature of our disk model is that the thermal and
secular instabilities can grow in its middle part even if the radiation
pressure is negligibly small in the condition $\delta > 2/3$. 
In the realistic disk
systems, it would be difficult to maintain the steady mass accretion state
unless the $Pm$-dependence of MRI-driven turbulence is relatively
weak. 
\end{abstract}

\keywords{accretion, accretion disks --- instabilities --- magnetic
fields --- MHD}

\section{Introduction}\label{intro}
The physical mechanism to transport angular momentum in accretion disks is an 
important unsettled issue in astrophysics. It has been extensively studied for 
decades using theoretical and numerical procedures. The main purpose of
these studies is to find the efficient transport process of the angular
momentum to account for the powerful mass accretion deeply associated
with the release of the gravitational energy 
observed in astrophysical disk systems \citep{2007MNRAS.376.1740K}. 

\cite{1973A&A....24..337S} propose a pioneering disk model for describing the steady 
mass accretion, in which the efficiency of the angular momentum transport is 
parametrized and represented by a dimensionless viscosity parameter $ \alpha  = -
t_{r\phi}/p$, where $t_{r\phi}$ is the $r\phi$-component of the stress tensor, 
and $p$ is the total pressure. The phenomenological parameter $\alpha$
introduced in the disk model should be determined experimentally. This disk
model is widely accepted today as the 
standard, and referred to as "Standard Shakura-Sunyaev Model" 
\citep[][hereafter, we simply call it the '$\alpha$-model']{1998bhad.conf.....K}.

It is pointed out by \cite{2007MNRAS.376.1740K} from the observational
point of view that the viscosity parameter $\alpha$ should be at the
level $\alpha \simeq 0.1$ in 
order to explain the powerful outbursts from disk systems, such as dwarf 
novae, X-ray transients, and a fully ionized part of the disk in Active
Galactic Nucleus (AGNs). Since the required level for 
the viscosity parameter $\alpha$ is much larger than that resulting from the microscopic 
molecular viscosity, the turbulent viscosity becomes the most promising candidate for 
governing the angular momentum transport in astrophysical disk systems 
\citep[see also][]{1998ApJ...495..385H, 1999AcA....49..391S,
2001A&A...373..251D, 2004MNRAS.347...67S, 2004MNRAS.353..841L}.

The Keplerian disks are known to be linearly stable to the shear instability (Rayleigh 
instability) which could power the hydrodynamic turbulence. Even when the non-linear 
perturbations are imposed, the Keplerian disks can never reach the highly turbulent state 
in the Rayleigh stable condition \citep{1996ApJ...467...76B}. Numerical studies in 
addition show that the other hydrodynamic instabilities, such as convective and 
Papaloizou-Pringle instabilities, do not play a central role in inducing
the efficient angular momentum transport in the disk systems
\citep{1984MNRAS.208..721P, 1988ApJ...326..277B, 1996ApJ...464..364S,
1998RvMP...70....1B, 2006ApJ...636...63J}. 

\cite{1991ApJ...376..214B} focus, for the first time, on the magneto-rotational 
instability (MRI) which is originally found by \cite{1059JETP...36..995} and 
\cite{1960PNAS...46..253C} as a candidate for operating the magnetohydrodynamic (MHD) 
turbulence in the disk systems. It has essentially a local nature and destabilizes  
weakly magnetized differentially rotating systems with a negative shear rate $q \equiv
\mathrm{d}\ln\Omega/\mathrm{d}\ln r$, where $\Omega$ is the angular velocity and 
$r$ is the cylindrical radius. 
The non-linear MHD turbulence driven by the MRI has been broadly investigated as the 
leading mechanism for the turbulent angular momentum transport in disk systems 
using state-of-the-art numerical techniques and huge computer facilities
for the MHD simulations \citep{1992ApJ...400..595H, 1995Sci...269.1365H, 2004ApJ...605..321S, 
2007A&A...476.1123F}. The physical properties of the MRI-driven turbulence, however, 
have not yet been confirmed completely even today. 

A recent remarkable finding in the MRI studies is the dependence of the transport 
efficiency of the MRI-driven turbulence on the magnetic Prandtl number $Pm$
\citep{2007MNRAS.378.1471L, 2007A&A...476.1123F, 2009ApJ...707..833S}, where 
$Pm = \nu/\eta$, $\nu$ is the viscosity and $\eta$ is the magnetic diffusivity. 
The local shearing box simulations explicitly taking into account the microscopic 
kinematic viscosity and resistivity suggest that the 
total turbulent stress (sum of Maxwell and Reynolds stresses) depends on the magnetic 
Prandtl number. Especially in the system with the non-zero net 
magnetic flux, the turbulent stress and the viscosity parameter
$\alpha$ resulting from the 
MRI-driven turbulence would increase with the magnetic Prandtl number, that is 
$\alpha \propto Pm^{\delta}$, where $\delta$ is the power-law 
index with positive value \citep{2007A&A...476.1123F, 2009ApJ...707..833S}. 

It should be stressed that the viscous effect on the non-linear
MRI-turbulence is seemingly contradictory to that on the linear growth
of the MRI \citep{2008ApJ...689.1234M,
2008ApJ...684..498P}. Surprisingly, the viscosity can boost the
transport efficiency of the MRI-driven turbulence although it suppresses
the growth of the MRI in its linear evolutionary regime. The physical
mechanism for the viscous boosting of the MRI-driven turbulence is a
central issue in the current MRI studies and remains to be solved
to draw a complete physical picture of the angular momentum
transport in the accretion disks \citep{2010arXiv1004.1384L}.

The impact of the microscopic diffusivities on the MRI-driven turbulence should be 
essential not only in understanding the nature of the MRI itself, but also in 
applying it to the realistic astrophysical disk systems. Assuming the standard 
$\alpha$-disk, the size of the magnetic Prandtl number drastically changes depending 
on the disk radius. \cite{2008ApJ...674..408B} suggest that the magnetic Prandtl number 
$Pm$ is typically larger in the inner disk when we adopt Spitzer's values
for evaluating the microscopic diffusivities \citep{1962pfig.book.....S}. Hence, the 
transport efficiency of the angular momentum should be affected by the magnetic Prandtl 
number and become a function of the disk radius when the viscosity parameter follows 
the relation $\alpha \propto Pm^{\delta}$ which is implied from the recent 
local shearing box simulations \citep{2007MNRAS.378.1471L, 2009ApJ...707..833S}.

In this paper, we construct a steady model for geometrically thin disk taking account 
of the "non-standard" description of the viscosity parameter which depends on the magnetic 
Prandtl number in the form $\alpha \propto Pm^{\delta }$. This should be 
consistent treatment of the turbulent viscosity with the results from recent local 
shearing box simulations \citep{2007MNRAS.378.1471L, 2009ApJ...707..833S}. A central 
difference between the work done by \cite{2008ApJ...674..408B} and ours is to include 
the structural change of the disk depending on the size of the magnetic Prandtl number 
$Pm$ in our model consistently. 

This paper is organized as follows: The steady disk model with the non-standard 
description of the viscosity parameter $\alpha \propto Pm^{\delta}$ is 
constructed in \S~2. We then describe the basic properties of our 
disk model. In \S~3, we investigate the stability of our steady disk model to the 
gravitational, thermal, and secular instabilities. The dependence of the stability 
criteria on the magnetic Prandtl number (or index $\delta$) is our interest in this 
section. Finally, we discuss the application of our model to the realistic 
astrophysical disk system and summarize our findings in \S~4. 
\section{Geometrically Thin Disk with Non-Standard $\alpha$-prescription}
We construct a steady disk model consistently combining the properties of the 
MRI-driven turbulence recently found by numerical studies \citep{2007MNRAS.378.1471L,
2009ApJ...707..833S}. The fundamental assumption adopted in this work is that the 
viscosity parameter $\alpha $ depends on the disk radius and reflects the transport 
properties of the MRI-driven turbulence. 
Also we assumed that the disk heating is local in which the
variation of the thermal energy responds to the turbulent energy
dissipation instantaneously \citep{1999ApJ...521..650B}. This assumption
is consistent
with the recent numerical work of the MRI-driven turbulence
\citep{2001ApJ...561L.179S, 2009ApJ...690..974S, 2009ApJ...694.1010G}.  

According to the recent numerical studies, the efficiency of the angular momentum 
transport sustained by the MRI-driven turbulence is regulated by the magnetic Prandtl 
number which characterizes the system. Using the viscosity parameter $\alpha$, the 
transport efficiency of the MRI-driven turbulence can be represented by, as is
described in \S~1, 
\begin{equation}
 \alpha = \alpha_0Pm^{\delta}, \label{eq:alpha}
\end{equation}
\citep{2007MNRAS.378.1471L, 2009ApJ...707..833S}, where $\alpha_0$ is the normalization 
parameter which controls the size of $\alpha $ to be smaller than unity. Note that 
this relation reduces to the classical standard $\alpha$-model by taking $\delta = 0$. 

For providing the magnetic Prandtl number, we adopt the Spitzers' values for the 
microscopic kinematic viscosity $\nu$ and the electron resistivity
$\eta$ by assuming the fully ionized gas. This assumption is
reasonable for the disk of the X-ray binaries and for the inner part of
the disk in AGNs. Then $\nu$ and $\eta$ are expressed as
\begin{equation}
 \nu = 1.6\times 10^{-15}\rho^{-1}T^{\frac{5}{2}} (\ln \Lambda_{HH})^{-1} \;, 
\label{eq:viscosity}  
\end{equation}
and
\begin{equation}
 \eta = 5.55\times 10^{11}T^{-\frac{3}{2}} \ln  \Lambda_{eH} \;, \label{eq:resistivity} 
\end{equation}
where $\rho$ is the disk mass density, and $T$ is the disk temperature 
\citep{1962pfig.book.....S}. The Coulomb logarithms for proton-proton and 
electron-proton scatterings are represented above by $\ln\Lambda_{HH}$ and 
$\ln\Lambda_{eH}$, respectively. The magnetic Prandtl number is then
described  as
\begin{equation}
 Pm=\left(\frac{T}{4.2\times 10^4~\mathrm{K}}\right)^4 
  \left(\frac{10^{14}~\mathrm{cm}^{-3}}{n l}\right),\label{eq:Pm}
\end{equation}
where $n$ is the number density and $l = \ln\Lambda_{eH}\ln\Lambda_{HH}$ is the 
product of the Coulomb logarithms and fixed to be $l=40$ in the following 
\citep[see][]{2008ApJ...674..408B}.

The governing equations are the conventional used in the classical standard 
$\alpha$-model:
\begin{itemize}
 \item[-] Mass conservation equation
      \begin{equation}
       \dot M = -2\pi r v_r \Sigma \;. \label{eq:cont}
      \end{equation}
 \item[-] Force balancing equation (Keplerian rotation)
       \begin{equation}
	\Omega=\Omega_\mathrm{K}\equiv\sqrt{\frac{GM}{r^3}} \;.\label{eq:Kepler}
       \end{equation}
 \item[-] Angular momentum conservation
       \begin{equation}
	\nu_t\Sigma = \frac{\dot M}{3\pi}\left(1-\sqrt{\frac{R_\mathrm{in}}{r}} \right) 
\equiv \frac{\dot M}{3\pi} f \;. \label{eq:angular}
       \end{equation}
 \item[-] Energy balancing equation
      \begin{equation}
       \frac{9}{4}\nu_t \Sigma \Omega^2 = \frac{32\sigma T^4}{3\tau} \;. \label{eq:energy}
      \end{equation}
 \item[-] Hydrostatic balance in the vertical direction
      \begin{equation}
       H = \frac{c_s}{\Omega} \;. \label{eq:hydro}
      \end{equation}
\end{itemize}
where $\Omega_\mathrm{K}$ is the Keplerian angular velocity. The physical parameters $M$, 
$\dot{M}$, $v_r$, $c_s$, $\nu_t$, $H$, $R_{\rm in}$ and $\tau$ are the mass of 
the central objects, the mass accretion rate, the radial velocity, the sound speed, 
the turbulent viscosity, the vertical scale height of the disk, the inner radius of the 
disk, and the optical depth in the vertical direction. Here $\Sigma $ is the surface 
density defined by $\Sigma = 2\rho H $. The physical constants $G$, $c$, and $\sigma$ 
are the gravitational constant, the speed of light, and the Stefan-Boltzmann constant
respectively. 

In this paper, we naively assume that the viscosity parameter $\alpha $ is related 
to the turbulent viscosity by a relation,
\begin{equation}
 \alpha p =\frac{3}{2}\rho\nu_t\Omega \;,\label{eq:alphapre}
\end{equation}
\citep[but, see][and \S~\ref{discuss} in this
paper]{2004ApJ...605..321S, 2007ApJ...668L..51P}. 
Note that there is a clear distinction 
between the turbulent viscosity $\nu_t$ and the kinematic
viscosity $\nu$. 
In our model, the turbulent viscosity $\nu_t $ is determined by the
MRI-driven turbulence which controls the accretion dynamics. 
The transport efficiency due to the MRI-driven turbulence is regulated
via the relation (\ref{eq:alpha}) by the microscopic kinematic viscosity
$\nu$. The kinematic 
viscosity affects the global nature indirectly by regulating the MRI-driven 
turbulence.

The opacity of the disk $\kappa $, which relates to the optical depth $\tau$ as 
$\kappa = 2\tau/\Sigma $, is given by
\begin{eqnarray}
 \kappa & = & \kappa_\mathrm{es}+ \kappa_{ff} \nonumber \\
        & = & \kappa_\mathrm{es}+\kappa_0\rho T^{-\frac{7}{2}} \;. \label{eq:opacity}
\end{eqnarray}
where $\kappa_{\rm es} \simeq 0.4\ {\rm cm^2\ g^{-1}}$ is the opacity due to the 
electron scattering and $\kappa_{\rm ff} $ is that contributed from the free-free 
absorption with $\kappa_0 \simeq 6.4\times 10^{22}$ in $cgs$ unit.

For the closure of the system, we need the equation of state for the composite gas 
of the photon and baryon,
\begin{equation}
 p = \frac{\rho}{\mu m_p}{k_B T}+\frac{4\sigma T^4}{3c}.\label{eq:eos}
\end{equation}
where $m_p$ is the proton mass, $k_B$ is the Boltzmann 
constant and $\mu $ is the mean molecular weight providing $\mu = 0.5$ throughout 
this paper.
\begin{figure}
\includegraphics[width=8cm]{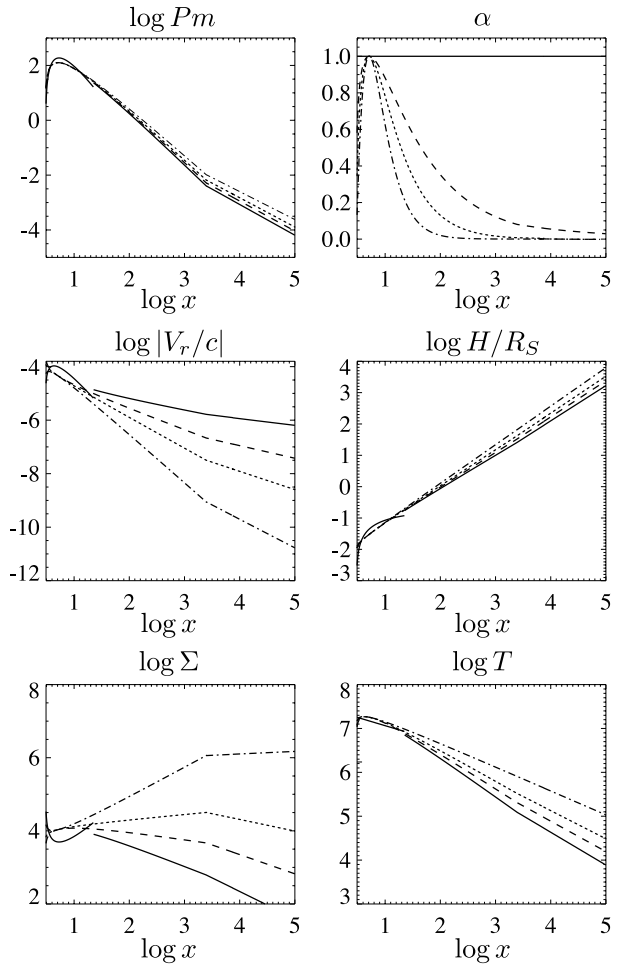}
\caption{Radial profiles of the magnetic Prandtl number (left top), $\alpha$
(right top), the radial velocity (left middle), 
the scale height (right middle), the surface density (left bottom),
and the temperature (right bottom).
$\alpha_0$ is determined so that the maximum of $\alpha$ is unity. 
Thick solid curves denote the solutions for the conventional standard disk
model, while dashed, dotted, and dash-dotted ones do those for
$\delta=0.25, 0.5, 1$, respectively. The other parameters are $m=10$,
$\dot{m}=1$.}
\label{fig:m1mdot0}
\end{figure}

Solving coupled equations (\ref{eq:alpha}), (\ref{eq:Pm})-(\ref{eq:eos})
with the zero stress boundary condition at $r = R_\mathrm{in}$ 
\cite[see, equation \ref{eq:angular}, or,][]{1973blho.conf..343N}, we can 
obtain the local disk structures for three characteristic regimes.
In the inner part of the disk, the radiation pressure dominates the gas
pressure. In contrast, the gas 
pressure becomes dominant in the middle and outer regions of the disk in our model. The 
difference between the middle and outer regions is the source responsible for the 
opacity. The electron scattering plays a crucial role in determining the opacity in 
the middle region. The outer structure of the disk depends mainly on the 
opacity contributed by the free-free absorption \citep[see, also standard 
$\alpha$-model in ][]{1973A&A....24..337S}.

The steady disk structure obtained here is then characterized by four dimensionless 
parameters, $x = r/r_s$, $m = M/M_\sun$, $\dot{m} = \dot{M}c^2/L_E$ and $\delta = 
\log(\alpha/\alpha_0)/\log Pm $, where $r_s$ is the Schwarzschild radius, 
$M_\sun$ is the solar mass, and $L_E$ is the Eddington luminosity, respectively. The 
innermost radius $R_\mathrm{in}$ is assumed as  $R_\mathrm{in}=3r_s$ in the following. 
\begin{figure}
\includegraphics[width=8cm]{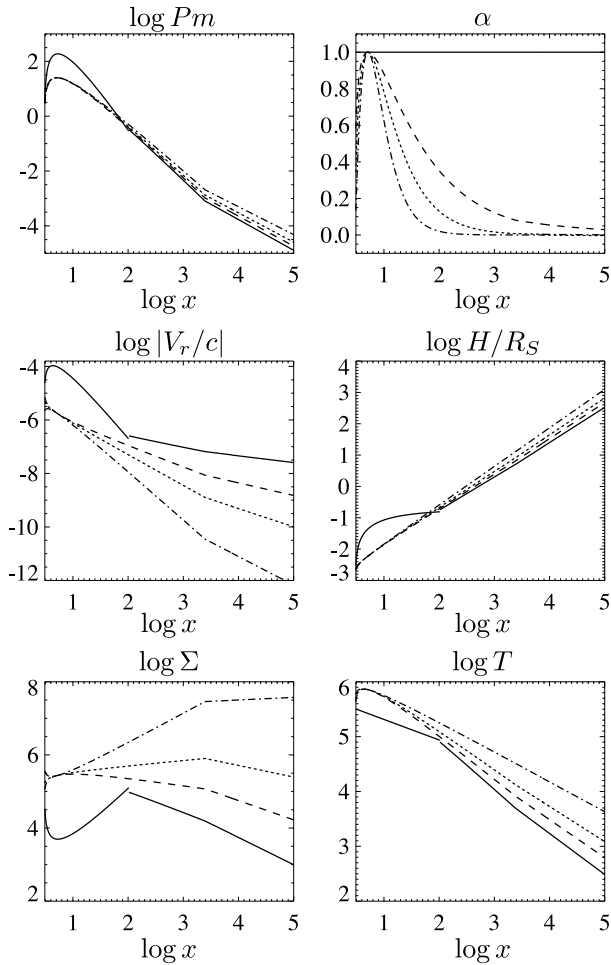}
\caption{Radial profiles of the magnetic Prandtl number (left top), $\alpha$
(right top), the radial velocity (left middle), 
the scale height (right middle), the surface density (left bottom),
and the temperature (right bottom).
$\alpha_0$ is determined so that the maximum of $\alpha$ is unity. 
Thick solid curves denote the solutions for the conventional standard disk
model ($\delta=0$), while dashed, dotted, and dash-dotted ones do those for
$\delta=0.25, 0.5, 1$, respectively. The other parameters are $m=10^8$,
$\dot{m}=1$.}
\label{fig:m8mdot0}
\end{figure}

Figures \ref{fig:m1mdot0} and \ref{fig:m8mdot0} show the radial profiles of the 
magnetic Prandtl number (left top), $\alpha$ (right top), the radial velocity in 
unit of $c$ (left middle), the scale height (right middle), the surface density 
(left bottom), and the disk temperature (right bottom) for the models
$m=10$ (Fig.~\ref{fig:m1mdot0}: 
the model X-ray binaries) and $m=10^8$ (Fig.~\ref{fig:m8mdot0}: the
model AGNs), respectively \citep[c.f.,][]{2008ApJ...674..408B}. The mass
accretion rate is fixed as the Eddington one, that 
is $\dot{m}=1$ in both figures. The normalization parameter $\alpha_0 $ is an 
arbitrary constant and is determined for the viscosity parameter $\alpha$ being unity 
at its maximum. Solid curves denote the solutions for $\delta = 0$ (conventional 
standard $\alpha$-model), while the dashed, dotted, and dash-dotted curves are for 
the cases $\delta=0.25, 0.5, 1.0$ respectively. 

We find that the magnetic Prandtl number is smaller than unity in the 
outer part of the disk, while it increases with decreasing the radius. Around the 
middle part of the disk, there is a critical radius $r_c$ where the magnetic Prandtl 
number exceeds unity. It exists around $r_c \simeq 100 r_s$ for both models. This is 
qualitatively consistent with the result obtained in
\cite{2008ApJ...674..408B}. 
The kinematic viscosity increases with the disk temperature and the
magnetic diffusivity decreases with it. Since the disk temperature is a
decreasing function of the disk radius, 
the magnetic Prandtl number varies sharply with the disk 
temperature and becomes lower and lower with the increase in the disk radius [see 
equation~(\ref{eq:resistivity})] except the inner disk. 
It would be important
to note that the magnetic Prandtl 
number is insensitive to the index $\delta$ although the disk temperature largely 
changes depending on it in the outer part of the disk. This is because the surface 
density varies more rapidly with the disk radius compared to the temperature not for 
violating the energy conservation law [see equation~(\ref{eq:energy})]. 

The efficiency of the angular momentum transport and the viscous heating, which 
can be represented by the viscosity parameter $\alpha$, must be lower in the outer 
part of the disk in our model. In addition, the parameter $\alpha$ strongly 
depends on the index $\delta$. We stress here that, for the model with the larger 
index $\delta$, the viscosity parameter $\alpha$ changes more steeply with radius. 
This tendency can be understood from equation~(\ref{eq:alpha}) by considering that 
magnetic Prandtl number is insensitive to the index $\delta$. The lower transport 
efficiency of the angular momentum results in the higher density and the lower accretion 
velocity in the outer part of the disk to ensure the constant mass accretion rate 
[see equation~(\ref{eq:cont})]. The drastic structural change in the middle
and outer regions of the disk 
is a natural consequence of the $Pm$-dependence of the viscosity parameter 
$\alpha$. 

We mention the physical behaviors of the inner disk. In our model, there exists 
a maximum of the parameter $\alpha $ at around the innermost radius $R_{\rm in}$. 
The disk structure is drastically changed across this point. 
Since we adopted the zero stress boundary condition at the inner radius
\citep{1973blho.conf..343N}, the gaseous material
cannot rotate with the Keplerian velocity inside $R_\mathrm{in}$. 
The gas would thus accrete to the central object while being less
affected by the turbulent viscous heating in such a region. This results
in the low disk temperature and then the low 
transport efficiency of the angular momentum (plunging region). This
should be the reason why there is an inflection point in the disk
structure in our model. 

\section{Stability of Disk Model}\label{sub:stability}
The inclusion of the $Pm$-dependence in the viscosity parameter $\alpha $ 
makes drastic changes in the disk structure. We investigate the linear stability 
of the disk model with "non-standard" $\alpha$-prescription to the gravitational, 
thermal, and secular instabilities. 
\subsection{Gravitational Instability}
As is shown in \S~2, the surface density in our disk model is larger
for a larger
$\delta$, especially in the outer region, than that in the standard 
$\alpha$-disk model. This is because the transport efficiency of the angular momentum 
becomes lower in the outer disk when we consider the $Pm$-dependence of the 
viscosity parameter. The larger amount of the gaseous material must be distributed 
there for the larger value of the index $\delta $. It would be thus affected more 
by the self-gravity in our disk model than the standard $\alpha$-model. 

We restrict our attention to the disk whose potential is dominated by the central 
object and whose rotation curve is therefore Keplerian. 
The gravitational instability to axisymmetric perturbations then sets in when the 
sound speed $c_s$, the rotation frequency $\Omega$, and the surface density $\Sigma $ 
satisfy
\begin{equation}
Q = \frac{\Omega c_s}{\pi G\Sigma} \lesssim 1 \;, \label{Toomore} 
\end{equation}
\citep{1964ApJ...139.1217T, 1965MNRAS.130..125G}. The instability condition 
(\ref{Toomore}) can be rewritten, for a disk with scale height $H\simeq c_s/\Omega$ 
around a central object of mass $M$,
\begin{equation}
M_{\rm disk} \gtrsim \frac{H}{r} M \;, 
\end{equation}
where $M_{\rm disk} = \pi r^2 \Sigma $ \citep[see][for details]{2003ApJ...597..131J}. 
The disk becomes unstable to the gravitational instability when the self-gravity of 
the gaseous material overcomes the gravitational force provided by the central object 
acting upon it. 

\begin{figure}
\includegraphics[width=8cm]{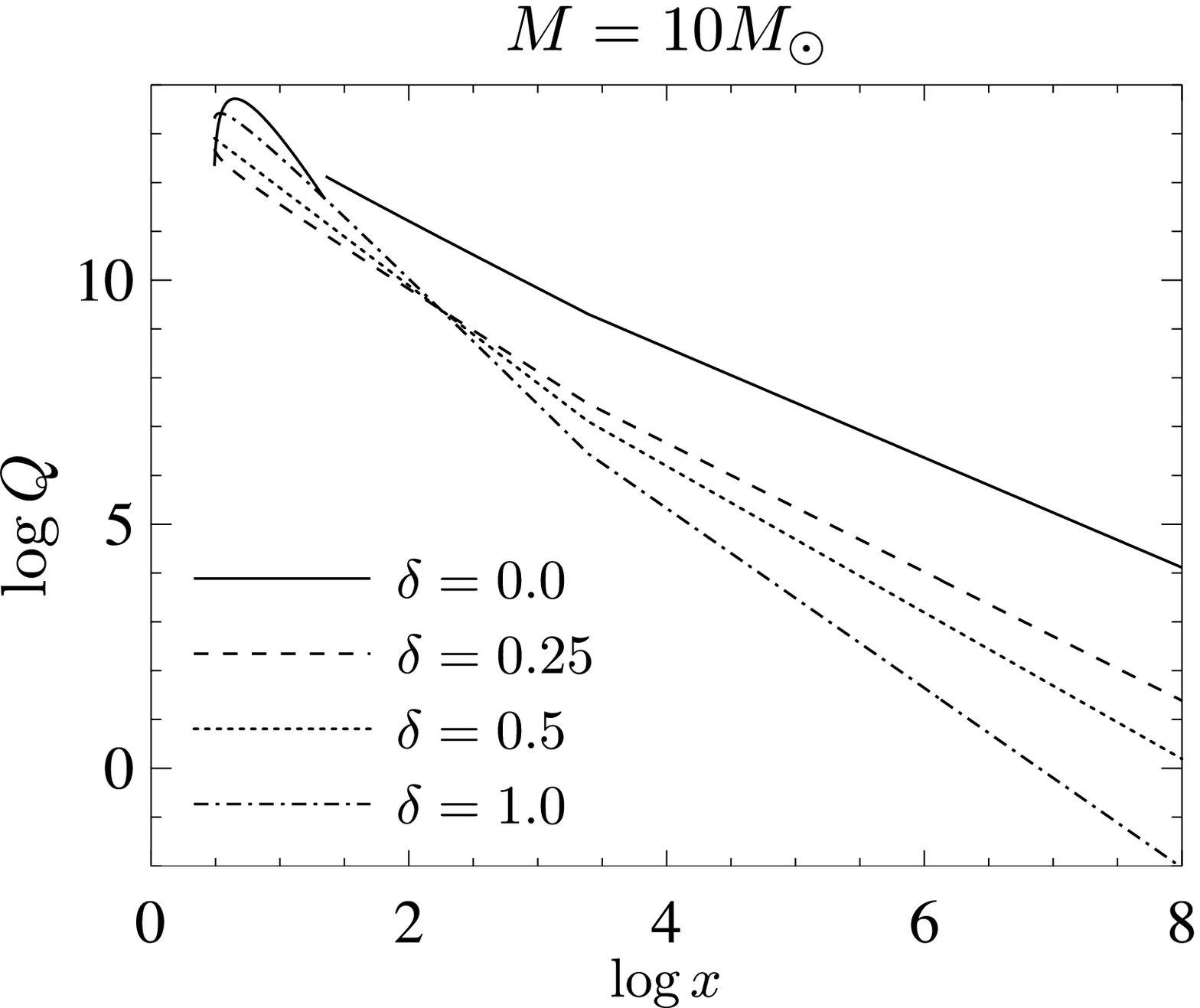}
\caption{Radial profile of Toomre's $Q$-value for $m=10$. Solid,
dashed, dotted, and dash-dotted curves represent for $\delta=0, 0.25,
0.5, 1.0$, respectively.}
\label{fig:critQM1}
\end{figure}

\begin{figure}
\includegraphics[width=8cm]{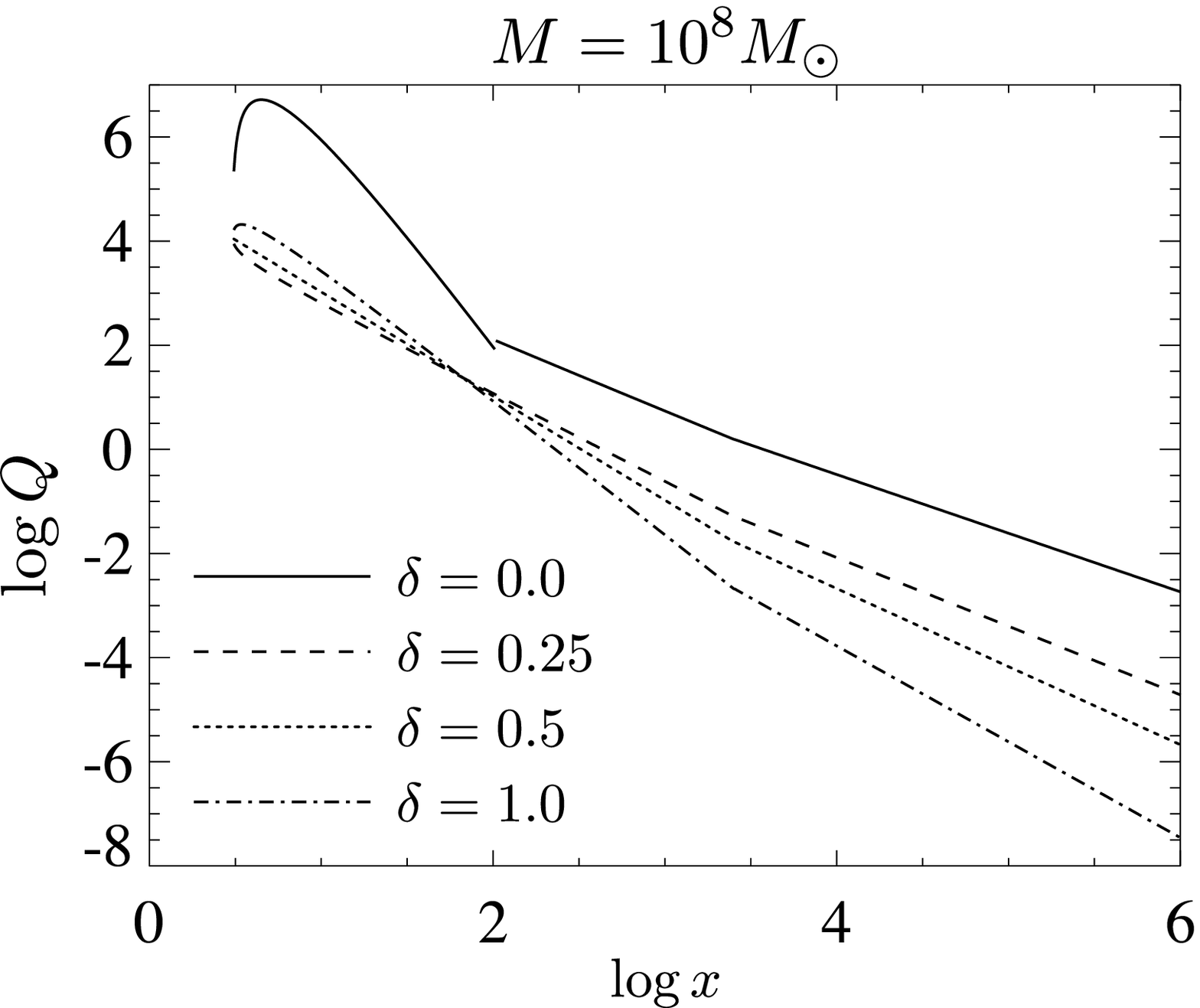}
\caption{Radial profile of Toomre's $Q$-value for $m=10^8$. Solid,
dashed, dotted, and dash-dotted curves represent for $\delta=0, 0.25,
0.5, 1.0$, respectively.}
\label{fig:critQM8}
\end{figure}

Figures \ref{fig:critQM1} and \ref{fig:critQM8} show the radial profiles of Toomre's 
$Q$-value for the models $m=10$ and $m=10^8$. Solid, dashed, dotted, and dash-dotted 
curves denote the cases $\delta = 0, 0.25, 0.5, 1.0$, respectively. The
mass accretion rate is fixed as $\dot{m}=1$ in both figures. 
It is found from 
these figures that Toomre's $Q$-value is smaller for the model with the larger index 
$\delta$. This is because the larger amount of gaseous material with
the slower 
accretion velocity should be in the outer regions for the larger index $\delta$. 
Hence, the disk becomes more unstable to the gravitational instability for the models 
with the larger index $\delta $. 

When $m=10$, Toomre's $Q$-value becomes smaller than unity in the outer part where 
$ r \gtrsim 10^8 r_s$. The X-ray binaries, which are the binary system of the stellar 
mass black hole and the companion stellar object, would be gravitationally stable 
because the typical size of these systems would be much smaller than $10^{8} r_s$. In 
contrast, for the model  $m=10^8$ (AGNs), Toomre's $Q$-value is smaller than unity 
in the region $r \gtrsim 10^2$--$10^3 r_s$. 
This indicates that the geometrically thin disk characterizing AGN systems is 
gravitationally unstable, not only in our model, but also in the conventional 
$\alpha$ model. The $Pm$-dependence of the viscosity parameter, which 
is the assumption based on the recent MRI studies, enhances the self-gravitational 
instability of the accretion disk because of the lower transport efficiency of the 
angular momentum in the region characterized by the lower magnetic Prandtl number. 

The self-gravity should play a significant role in enhancing the turbulent viscosity, 
and triggering the disk fragmentation \citep{1980ApJ...242..209B,1982ApJ...256..390S, 
1987MNRAS.225..607L, 1987Natur.329..810S}. Although we must take account of these 
self-gravity effects consistently in our governing equations to construct the disk 
model correctly capturing the AGNs disk system 
\citep{1979AcA....29..157K, 1996ApJ...469L..49M}, it is beyond the scope
of this paper. The main purpose of this 
paper is to extend the standard $\alpha$-disk to the model with $Pm$-
dependence of the viscosity parameter as the first step. 

\subsection{Thermal and Secular Instabilities}\label{TSins}
Equations~(\ref{eq:alpha}) and (\ref{eq:Pm}) indicate that the lower disk temperature 
provides the lower transport efficiency of the angular momentum, and therefore the 
lower viscous heating rate. The radial variations of the viscous parameter 
associated with radially declining viscous heating would impact on the responses of 
the disk model to the thermal and secular instabilities \citep{2008ApJ...674..408B}. 

Since the thermal timescale is much longer than the dynamical one, we can assume that 
the dynamical equilibrium is retained when we consider the thermal instability. In 
addition, we can assume that the surface density of the disk does not change in the 
thermal evolution time because the centrifugal force balances with gravitational force 
in the radial direction. Thus the gaseous motion caused by 
the temperature perturbation is restricted almost in the vertical 
direction. These assumptions can simplify the treatment of the thermal instability. 

We consider the thermal equilibrium as the unperturbed state in the condition with 
the fixed surface density. When the disk temperature is slightly perturbed over 
that of the equilibrium state, the criterion for the thermal instability is given 
\begin{equation}
\left. \frac{\partial (Q_\mathrm{vis}^+ - Q_\mathrm{rad}^-)}{\partial
 T}\right|_{\Sigma = \mathrm{const}}> 0 \;,\label{crit:thermal}
\end{equation} 
where 
\begin{eqnarray}
Q_\mathrm{vis}^+ & = & -\frac{3}{2}\Omega T_{r\phi} \;, \label{eq:Qvis}\\
Q_{\rm rad}^- & = & \frac{32\sigma T^4}{3\tau}  \;, \label{eq:Qrad}
\end{eqnarray}
\citep{1973A&A....24..337S, 1976MNRAS.177...65P,1978ApJ...221..652P}. Here 
$Q_\mathrm{vis}^+$ is the viscous heating rate \citep{1976MNRAS.177...65P}, and 
$Q_{\rm rad}^- $ is the radiative cooling rate. The thermal equilibrium satisfies the 
condition $Q_\mathrm{vis}^+ = Q_\mathrm{rad}^-$. Note that the viscous heating and 
radiative cooling rates are both the increasing function of the disk temperature. 
The disk becomes thermally unstable when the increasing rate of the viscous 
heating overcomes that of the radiative cooling. Once the disk temperature is 
slightly increased from the equilibrium state, it is further enhanced by the positive 
feedback from the viscous heating which is more effective than the radiative cooling 
\citep{1998bhad.conf.....K}.

The secular instability is a phenomenon resulting from the spatial modulation of the 
accretion rate. The typical timescale in which the instability grows is thus the 
viscous timescale, which is much longer than the thermal and dynamical ones in the 
geometrically thin disk. Thus we can assume that the state is in thermal
and dynamical equilibria, so that the thermal 
and oscillatory modes can be filtered out. Then the criterion for the secular 
instability is
\begin{equation}
 -\left.\frac{\partial T_{r\phi}}{\partial
   \Sigma}\right|_{Q_\mathrm{vis}^+=Q_\mathrm{rad}^-} < 0 \;,
 \label{crit:secular}
\end{equation}
\citep{1974ApJ...187L...1L, 1998bhad.conf.....K}
where $T_{r\phi} = -2\alpha p H$ is the $r\phi$-component of the turbulent stress 
tensor. 
This means that the instability sets in when the turbulent stress
decreases with increasing the surface density. 
Once the surface density is perturbed positively from the dynamical
equilibrium state, the turbulent stress and the efficiency of the angular
momentum transport decreases.
The perturbation of the surface density is thus 
further enhanced by the positive feedback ($\sim$ negative diffusion). 

For the simple treatment of two types of instabilities above, we introduce 
dimensionless parameters $\beta$ and $\gamma$. The ratio of the gas pressure 
$p_{\rm gas}$ to the total pressure $p$, which is the sum of the gas 
pressure and the radiation pressure $p_{\rm rad}$, can provide the
parameter $\beta$ \citep{1978ApJ...221..652P}
\begin{equation}
 \beta \equiv \frac{p_\mathrm{gas}}{p} \;.\label{eq:beta}
\end{equation} 
The $r$-$\phi$ component of the stress tensor $T_{r\phi}$ has a general form
 \begin{equation}
 T_{r\phi} = -2 \alpha p_\mathrm{gas}^\gamma p^{1-\gamma} H \;. \label{eq:galphapre}
\end{equation}
From these definition, the parameters $\beta$ and $\gamma$ range
$0 \le \beta, \gamma \le 1$.
The solution of $\gamma = 0$ corresponds to that of the Shakura-Sunyaev disk, and 
$\gamma = 1$ to that where $T_{r\phi}$ is proportional to the gas pressure instead of 
the total pressure.

By adding the small-amplitude perturbations to the systems described by set
of equations~(\ref{eq:Pm})--(\ref{eq:eos}) including the
$Pm$-dependence of the viscosity parameter $\alpha$ with the assumptions
discussed above, we obtain the criteria for the thermal and secular
instabilities with the parameters $\delta$,
$\beta$, and $\gamma$, (see, appendix for derivation).

When we assume the electron scattering mainly contributes to the
opacity, we can combine the criteria for the thermal and secular
instabilities to a single condition \citep{1998bhad.conf.....K} as
\begin{equation}
4-10\beta-7\gamma(1-\beta) +\delta(8+\beta) > 0. \;\label{tins:mid}
\end{equation}
 Note that the $\alpha$-prescription depending 
on $Pm$ is adopted here instead of the conventional approach. 
Neglecting the last term resulting from the $Pm$-dependence of the viscous 
parameter, equation (\ref{tins:mid}) reduces to the criterion of thermal and secular 
instabilities for the standard $\alpha$-model. Since the last term is always positive
for the case $\delta \geq 0$, 
our disk model with 'non-standard' $\alpha$-parameter 
should be more destabilized when the transport efficiency of the angular momentum by 
the MRI-driven turbulence depends more strongly on the magnetic Prandtl number. 

\begin{figure}
\includegraphics[width=8cm]{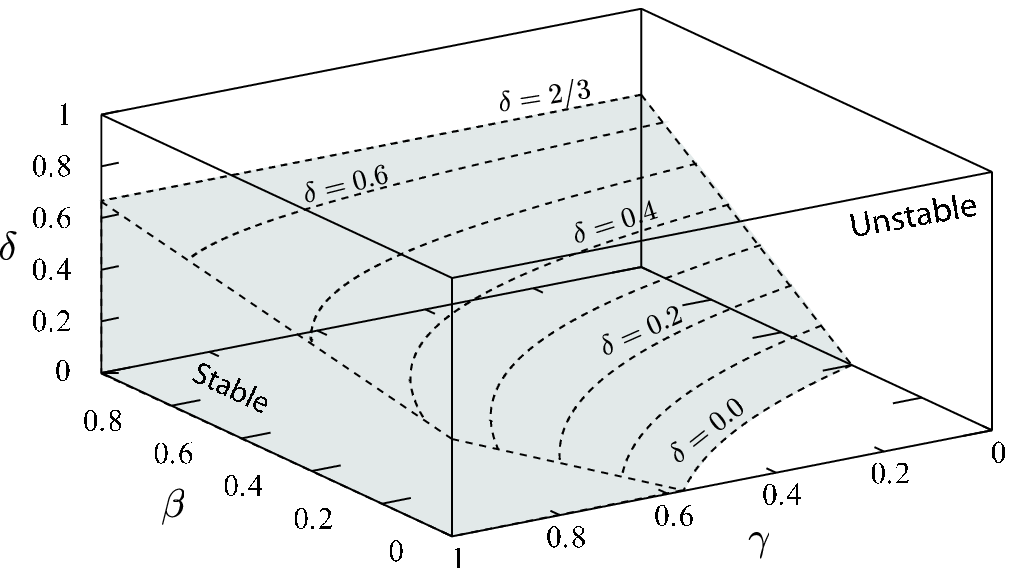}
\caption{Thermal and secular unstable domain on the parameter space of
$\beta-\gamma-\delta$. We assume electron-electron
scattering for the main opacity. The shaded domain 
shows the stable region.}
\label{fig:critmiddle}
\end{figure}

Figure \ref{fig:critmiddle} shows the stability criterion for the thermal and secular 
instabilities in the parameter space of $\beta$--$\gamma$--$\delta$.
The shaded domain indicates the stable region. The disk is unstable for the model 
with the smaller $\beta$ and $\gamma$. This indicates that the instabilities are 
generally facilitated when the contribution of the radiation pressure to the total 
pressure becomes larger. 
It is characteristic of our disk model that the unstable domain extends
for a larger $\delta$,
indicating that the disk is more unstable when the magnetic Prandtl
number strongly affects the angular momentum transport in MRI-driven
turbulence. 
\begin{figure}
\includegraphics[width=8cm]{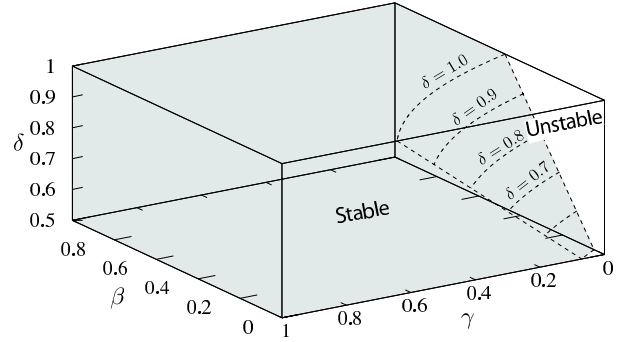}
\caption{Thermal and secular unstable domain on the parameter space of
$\beta-\gamma-\delta$. We assume free-free opacity.
The shaded domain shows the stable region.} 
\label{fig:critout}
\end{figure}

Considering the Shakura-Sunyaev type stress tenor ($\gamma = 0$), the condition for 
the instability is rewritten by
\begin{equation}
 \beta < \frac{4 + 8\delta}{10 - \delta} \;.
\end{equation}
The disk becomes unstable when $\beta < 0.4$ in the condition $\delta=0$ (standard 
$\alpha$-model). The unstable parameter space extends when $\delta > 0$. This can be 
understood intuitively from equations (\ref{eq:alpha}) and (\ref{eq:Pm}). 
Our model gives the relation $\alpha \propto \rho^{-\delta}T^{4\delta}$ and thus the 
viscous heating rate increases with the disk temperature when $\delta > 0$. 
Since the larger index $\delta$ provides more efficient viscous heating, it should
play a significant role in enhancing the thermal instability. On the other hand, the 
transport efficiency of the angular momentum decreases with the density when 
$\delta > 0$. The positive feedback effect becomes larger for the model with the 
larger $\delta$. Hence the increase of the index $\delta $ also enhances the secular 
instability. 

We would like to stress here that the disk becomes unstable even if the radiation 
pressure is negligibly small ($\beta = 1$) when the condition $ \delta > 2/3$ 
is satisfied. Recent local shearing box simulations for the MRI imply that the index 
$\delta$ ranges from $0.25$ to unity \citep{2007MNRAS.378.1471L, 2009ApJ...707..833S}.  
This suggests that the accretion disk sustained by the MRI-driven turbulence can become 
thermally and secularly unstable even when the gas pressure dominates the radiation 
pressure ($p_{\rm gas} \gg p_{\rm rad}$). This is an important property
that does not appear in the standard $\alpha$-model. 

When the free-free emission is the dominant source of the opacity, the criteria for 
the thermal and secular instabilities are combined again to a single
condition (see appendix for derivation)
\begin{equation}
 7 + 21\beta + 14\gamma(1 - \beta) - 2\delta(8 + \beta) < 0 \;. \label{tins:out}
\end{equation}
Note that all terms except the last one take always positive. This suggests that 
the disk is always stable to the thermal and secular instabilities in the condition 
$\delta =0$ when the free-free opacity becomes dominant. In contrast, the last 
term is always negative when $\delta > 0$. The outer region of our model 
then tends to be unstable to these instabilities even when the opacity of the
disk is mainly contributed from the free-free absorption. 

Figure~\ref{fig:critout} demonstrates the stability criterion for
the thermal and secular instabilities in the parameter space of
$\beta$--$\gamma$--$\delta$. The shaded region indicates the stable
domain.
It is found that the parameter space for 
the disk being unstable to these instabilities extends with increasing the index 
$\delta$. It is important that the disk becomes unstable to these instabilities 
in the case $\delta > 0$ although it is thermally and secularly stable in the whole 
parameter domain when $\delta = 0$. 

The Shakura-Sunyaev type stress tensor ($\gamma=0$) reduces the 
condition~(\ref{tins:out}) to
\begin{equation}
 \beta <- \frac{7 - 16\delta}{21 - 2\delta} \;.\label{tins:out_g0}
\end{equation}
The conventional $\alpha$-disk ($\delta = 0$) is always stable
because $\beta $ should be positive.
In the case 
$\delta =0$, the radiative cooling rate $Q_\mathrm{rad}^-$ more steeply responds to the 
disk temperature than the viscous heating rate $Q_\mathrm{vis}^+$. The thermal 
instability is thus suppressed by the negative feedback resulting from the strong 
radiative cooling. When the viscosity parameter $\alpha$ depends on the magnetic 
Prandtl number, the viscous heating rate also becomes the function of the disk radius. 
Then the positive feedback effect is amplified and the disk becomes unstable because 
the viscous heating rate increases with the disk temperature more steeply
than the radiative cooling rate.  
Since the denominator of equation (\ref{tins:out_g0}) has a
positive value according to the recent local shearing box simulations,
the condition (\ref{tins:out_g0}) reduces to $\beta < 16\delta-7$.
Then the disk can be thermally and secularly unstable in the case
$\delta > 7/16$ when the radiation pressure dominates the gas pressure
($\beta = 0$) with Shakura-Sunyaev type stress tensor ($\gamma=0$). 

When the gas pressure dominates the radiation
pressure ($\beta = 1$), the criterion for the thermal and secular
instabilities is given as $\delta > 14/9$ from equation (\ref{tins:out}).
Thus the outer part of the accretion disks can be unstable to these
instabilities when the efficiency of the angular momentum
transport and the viscous heating rate are strongly controlled by the
magnetic Prandtl number.

\section{Summary and Discussion}\label{discuss}
We construct a disk model with the non-standard $\alpha$-prescription depending on the 
magnetic Prandtl number $\alpha \propto Pm^\delta $ according to the recent numerical 
studies for the MRI. The magnetic Prandtl number $Pm$ evaluated from the 
Spitzer's value is a decreasing function with disk radius and it becomes smaller than 
unity for the region $r \gtrsim 100 r_s$ \citep{2008ApJ...674..408B}.

As a result of $Pm$-dependence of the viscosity parameter $\alpha$, the transport 
efficiency of the angular momentum becomes lower in the outer part of the disk in the 
model with larger value of the index $\delta$. Since the accretion velocity 
decreases with $\delta$ due to the inefficient angular momentum
transport, there should be a large 
amount of gaseous materials in the outer part of the disk. The self-gravity enhanced 
as a result of the $Pm$-dependence of $\alpha$-parameter would make the disk more unstable 
to the self-gravitational instability especially for the AGN disks in the region 
$r\gtrsim 10^2-10^3 r_s$  \citep[see][for the case with standard $\alpha$-prescription]
{1978AcA....28...91P, 1979AcA....29..157K, 1992PASJ...44..553F, 1996ApJ...469L..49M}.

Since our model gives the relation $\alpha \propto \rho^{-\delta} T^{4 \delta}$, the 
viscous heating rate increases with the disk temperature when $\delta > 0$. This 
suggests that the positive temperature perturbation enhances the viscous heating rate 
and thus the positive feedback to the thermal instability. On the other hand, the 
transport efficiency of the angular momentum decreases with increasing
the gas density when 
$\delta > 0$. This works in the same way as the negative diffusion and
also enhances the positive feedback to the secular instability. 

It is the most remarkable feature of our disk model that the thermal and
secular instabilities can grow in its middle part even if the radiation
pressure is negligibly small in the condition $\delta > 2/3$. Also
the outer part of the disk can be unstable to thermal and secular
instabilities when $\delta>7/16$ (radiation pressure dominant with
Shakura-Sunyaev type stress tensor, $\gamma=0$) or
$\delta > 14/9$ (gas pressure dominant).
If the MRI plays an important role in operating 
the MHD-turbulence and the viscosity parameter $\alpha$ depends on the 
magnetic Prandtl number, the steady mass accretion would not be maintained due to 
the growth of these instabilities in the geometrically thin disk. The rapid 
transition from the steady accretion phase to the non-steady state should be 
considered to enable us to understand more deeply the origin of the
light curve observed from the realistic astrophysical disk system.

In our model, we assumed that the turbulent stress tensor is just
proportional to the pressure according to the standard $\alpha$-disk
model (see, equation \ref{eq:galphapre}). On the other hand, recent
local shearing box simulations for the MRI imply that the pressure
dependence of the turbulent viscosity might be weak compared to that
assumed in the standard $\alpha$-disk model, that is $T_{r,\phi}\propto
p^\epsilon$ where the index $\epsilon$ takes a value in the range
$1/6 \lesssim \epsilon \lesssim1/4$ \citep{2004ApJ...605..321S,
2007ApJ...668L..51P}. 


When the pressure dependence of the turbulent viscosity is weaker than
that assumed in the standard $\alpha$-disk, the viscous heating rate is
expected to become insensitive to the temperature perturbation. The disk
thus becomes more stable to the thermal instability. Additionally, the
secular instability should also be suppressed because the turbulent stress
would then respond weakly to the density perturbation. 


Now let us derive the stability condition of disks when the turbulent stress
varies weakly depending on the pressure as is implied from the recent
numerical studies. Using the prescription $T_{r\phi}= - 2
\alpha p_\mathrm{gas}^{\gamma} p^\epsilon H$ [note that this equation
reduces equation (\ref{eq:galphapre}) when $\epsilon = 1-\gamma$] and
assuming the electron
scattering as the dominant opacity source, the instability conditions are
given by (\ref{ap:midth}) for the thermal instability, and
(\ref{ap:midsc}) for the secular instability (see Appendix for the
details of derivation).
Conditions (\ref{ap:midth}) and (\ref{ap:midsc}) can commonly yield, by
taking $\beta=1$ (gas pressure dominant),
\begin{equation}
 \delta > \frac{7 - \gamma-\epsilon}{9},\label{sol:pthmid}
\end{equation}
Here we assume $\delta \geq 0$. This equation indicates that the
weak dependence of the viscosity parameter on the pressure (i.e.,
$|\gamma|, |\epsilon| \ll 1$) has a
stabilization effect on the thermal and secular instabilities. 

For the model in which the free-free absorption is the main opacity
source, the stability conditions for the thermal and secular
instabilities can be obtained from (\ref{ap:outth}) and
(\ref{ap:outsc}), respectively. By taking $\beta=1$,
these conditions are combined to a single condition 
\begin{equation}
 \delta > \frac{15-\gamma-\epsilon}{9}. \label{sol:pthout}
\end{equation}
Here we assume $\delta \geq 0$.
This also shows that the instabilities can be suppressed when the
pressure dependence of the viscosity parameter is weaker than that
assumed in the standard disk models. 

It should be important to stress that even when the turbulent stress
does not directly depend on the pressure (i.e., $\gamma=\epsilon = 0$), the disk can be
unstable since the efficiency of the angular momentum transport $\alpha$
depends on the temperature and the density. In such case, the
instabilities grow when $\delta > 7/9$ for the middle part of the disks
and when $\delta > 5/3$ for the outer part of the disks. 
Since $\delta$ is predicted to be less than unity from the recent local
simulations, the middle part of the disks can be unstable to these
instabilities when the saturation levels of MRI-driven turbulence is
strongly controlled by the magnetic Prandtl number.

Finally we discuss the validity of the assumption adopted in our disk
model. Numerical 
studies for the MRI imply that transport properties of the MRI-driven
turbulence depend on the magnetic Prandtl number as is given in equation
(\ref{eq:alpha}). 

This relation is verified only in the narrow parameter range 
$10^{-2}\lesssim Pm \lesssim 10$ and $10 \lesssim R_\mathrm{M} \lesssim
10^4$ \citep{2007MNRAS.378.1471L, 2010arXiv1004.1384L}. Here
$R_\mathrm{M}=Hc_s/\eta$ is the magnetic Reynolds number. 
The range 
of $Pm$ examined in numerical studies corresponds to the region 
$10 r_s \lesssim r \lesssim 10^3 r_s$ in our model. The $Pm$-dependence of the 
viscosity parameter $\alpha $ is not well studied due to the
computational issue in the parameter range characterizing 
the outer part of the disk model. 
In addition, Reynolds and magnetic Reynolds numbers are taken to be
involuntarily larger values than the realistic values for the
astrophysical ionized gas due to the limitation of the computational
resources. If we suppose the realistic values of viscosity and
resistivity, their related dissipation scales become much smaller than
those resolved in existing state-of-the-art numerical simulations. 
In order to facilitate the understanding of astrophysical accretion
processes, we need to examine whether such small scale dissipation
processes actually affect the dynamics of the accretion disks via the
angular momentum transport and then the instabilities of the disk. 

Additionally, it has not been confirmed whether the viscosity parameter
is impacted by the magnetic Prandtl number when the microscopic
diffusivities are the functions of the physical parameters such as the
density and the temperature because the uniform diffusivities are
assumed in the existing local shearing box simulations
\citep{2007MNRAS.378.1471L, 2007A&A...476.1123F, 2009ApJ...707..833S}.
We need further systematic studies on the non-linear properties of 
the MRI in the vertically stratified disks with variable diffusive
parameters and the realistic radiative cooling in order to verify
the $Pm$-dependence of the turbulent viscosity parameter $\alpha$.

Our model does not include just for simplicity the effect of
magnetic fields explicitly (but takes account of the effect of the MHD
turbulence through the non-standard 
$\alpha$-prescription). The impacts of the magnetic field can modify the disk structure. 
Since the magnetic pressure contributes to the total pressure appeared in equation 
(\ref{eq:galphapre}) and weakens the temperature dependence of the viscous heating, 
it might suppress the growth of gravitational, thermal and secular instabilities 
when the magnetic pressure dominates the gas and radiation pressures. This is beyond 
the scope of this paper but would be studied as our future work using numerical 
procedures.

\acknowledgments

We are grateful to an anonymous referee for improving our manuscript.
We also thank Hiroshi Oda, Ken Ohsuga and Ryoji Matsumoto for fruitful discussions. 

\appendix
\section{Criteria for the instabilities}\label{ap:derive}
In this appendix, we briefly summarize the derivation of the stability
conditions shown by equations (\ref{tins:mid}), (\ref{tins:out}),
(\ref{sol:pthmid}), and (\ref{sol:pthout}).
Here we use a following general prescription for the stress tensor,
instead of the form introduced by equation (\ref{eq:galphapre}),
\begin{equation}
T_{r\phi} = -2 \alpha p_\mathrm{gas}^\gamma p^{\epsilon} H \;, \label{ap:galpha}
\end{equation}
where $\epsilon$ is a constant. When $\epsilon = 1-\gamma$, this equation
reduces to equation (\ref{eq:galphapre}). We hereafter describe the unperturbed
and perturbed quantities by subscripts '0' and '1', respectively. Note
that all the physical unperturbed quantities are assumed to be much
larger than their associated perturbed quantities in the following. 

We consider the stability criterion of the system to the thermal
instability. As is discussed in \S~\ref{TSins}, we can assume that the
dynamical equilibrium is maintained during the growth of the
thermal instability since
the dynamical time scale is much shorter than the thermal one. Also we
can assume that the material motion resulting from a temperature
variation occurs only in vertical direction since the force balances
with the gravitational force in the radial direction. Then the surface
density $\Sigma$ is approximately unchanged during the growth of the
thermal instability;
\begin{equation}
 \Sigma = \mathrm{const},\label{ap:sconst}
\end{equation}
while the hydrostatic balance is maintained \citep{1998bhad.conf.....K}.  

The linearized form of equation (\ref{eq:hydro}) and the total
pressure $p=p_\mathrm{gas} + p_\mathrm{rad}$, where $p_\mathrm{rad}$ is
the radiation pressure, are 
\begin{equation}
 \frac{H_1}{H_0} = -\frac{1-\beta}{1+\beta}\frac{\Sigma_1}{\Sigma_0}
  + \frac{4-3\beta}{1+\beta}\frac{T_1}{T_0}, \label{ap:th1}
\end{equation}
\begin{equation}
\frac{p_1}{p_0}= \frac{2\beta}{1+\beta}\frac{\Sigma_1}{\Sigma_0}
 + \frac{4-3\beta}{1+\beta}\frac{T_1}{T_0}. \label{ap:th2}
\end{equation}
Note that equations
(\ref{eq:eos}) and (\ref{eq:beta}) are adopted for the derivation.
 
Then the perturbative equation of the viscous heating
(\ref{eq:Qvis}) can be obtained from (\ref{eq:alpha}), (\ref{eq:Pm}),
(\ref{ap:galpha}), and (\ref{ap:th1})-(\ref{ap:th2}) as
\begin{equation}
\frac{Q_{\mathrm{vis},1}^+}{Q_{\mathrm{vis},0}^+}
=\frac{(4-3\beta)(1+\epsilon)-\gamma(3-4\beta)+\delta(8+\beta)}
{1+\beta}
\frac{T_1}{T_0}
+\frac{-1+\beta-2\delta+2\gamma+2\beta\epsilon}{1+\beta}
\frac{\Sigma_1}{\Sigma_0},
\label{ap:th3}
\end{equation}

When the electron scattering is the dominant source of the opacity, the radiative cooling (\ref{eq:Qrad}) provides a linearized equation
\begin{equation}
 \frac{Q_{\mathrm{rad},1}^-}{Q_{\mathrm{rad},0}^-}=4\frac{T_1}{T_0} - \frac{\Sigma_1}{\Sigma_0}.\label{ap:th4}
\end{equation}
By substituting equations (\ref{ap:th3}) and (\ref{ap:th4}) into
equation (\ref{crit:thermal}) and using equation (\ref{ap:sconst}), we
obtain the instability condition
\begin{equation}
 \frac{4-10\beta-7\gamma(1-\beta) +\delta(8+\beta)-(4-3\beta)(1
  -\gamma-\epsilon)}{1+\beta} > 0. \label{ap:midth}
\end{equation}
When we take $\epsilon = 1 - \gamma$, this equation reduces to equation
(\ref{tins:mid}) since $\beta \ge 0$.

When the free-free absorption mainly determines the opacity of the system, we can give a linearized equation for the radiative cooling,
by using equations (\ref{eq:opacity}) and (\ref{ap:th1}),
\begin{equation}
 \frac{Q_{\mathrm{rad},1}^-}{Q_{\mathrm{rad},0}^-}
=\frac{23+9\beta}{2(1+\beta)}\frac{T_1}{T_0}
-\frac{3+\beta}{1+\beta}\frac{\Sigma_1}{\Sigma_0}. \label{ap:th6}
\end{equation}
By substituting equations (\ref{ap:th3}) and (\ref{ap:th6}) into equation (\ref{crit:thermal}) and applying equation (\ref{ap:sconst}), we obtain the instability criterion for the thermal instability,
\begin{equation}
 \frac{7+21\beta+14\gamma(1-\beta)-2\delta(8+\beta)+2(4-3\beta)(1-\gamma-\epsilon)}{2(1+\beta)}<0.
  \label{ap:outth}
\end{equation}
Note that this can reduce to the condition (\ref{tins:out}) when
$\epsilon=1-\gamma$ since $\beta \ge 0$.

Finally we consider the stability condition of the system against the secular instability. When the secular instability grows, the negative diffusion amplifies the surface density so that the assumption (\ref{ap:sconst}) cannot be adopted here.
As mentioned in \S~\ref{TSins}, the viscous time scale, which is equivalent with the typical growth time of the secular instability, is much longer than the thermal time scale. The thermal equilibrium is thus maintained during the development of the secular instabilities, i.e.,
$Q_{\mathrm{vis}}^+=Q_{\mathrm{rad}}^-$. This can provide a linearized equation
\begin{equation}
 (-4+\gamma+4\delta)\frac{T_1}{T_0}
+ (1+\gamma-\delta)\frac{\Sigma_1}{\Sigma_0}
+ (1-\gamma+\delta)\frac{H_1}{H_0}
+ \epsilon\frac{p_1}{p_0}=0,\label{ap:th7}
\end{equation}
for the case with the electron scattering as the main opacity source, and
\begin{equation}
 (-15+2\gamma+8\delta)\frac{T_1}{T_0}
  + (4+2\gamma-2\delta)\frac{\Sigma_1}{\Sigma_0}
  + 2(-\gamma+\delta)\frac{H_1}{H_0}
  + 2\epsilon\frac{p_1}{p_0}=0,\label{ap:th8}
\end{equation}
for the model with the free-free absorption as the dominant opacity source.

By combining equations (\ref{ap:th1}), (\ref{ap:th2}) and (\ref{ap:th7}) or (\ref{ap:th8}), we can express the quantities $H_1/H_0$, $T_1/T_0$ and $p_1/p_0$ by $\Sigma_1/\Sigma_0$ for each
model with different opacity source. Then by substituting these obtained
equations into equation (\ref{crit:secular}), we obtain the instability
criterion for the secular instability.

When the electron scattering is the main opacity source, the instability
condition is given as
\begin{equation}
 \frac{5\gamma+4\epsilon+\beta(1+4\gamma+5\epsilon+\delta)}
{4-10\beta-7\gamma(1-\beta)+\delta(8+\beta)-(4-3\beta)(1-\gamma-\epsilon)}>0.\label{ap:midsc}
\end{equation}
This condition is identical with equation (\ref{ap:midth}) when
$\delta>0$ since the numerator in equation (\ref{ap:midsc}) has a
positive value.
Especially when $\epsilon=1-\gamma$, this equation reduces to equation
(\ref{tins:mid}).

The disks in which the free-free absorption becomes the dominant opacity source is unstable to the secular instability when
\begin{equation}
\frac{1+28\gamma+24\epsilon+2\delta+\beta(3+8\gamma+12\epsilon+2\delta)}
{7+21\beta+14\gamma(1-\beta)-2\delta(8+\beta)+2(4-3\beta)(1-\gamma-\epsilon)}<0.\label{ap:outsc}
\end{equation}
Since the numerator of equation (\ref{ap:outsc}) is positive when
$\delta>0$, the condition for the secular instability in the outer
region is identical with that for the thermal instability given in
equation (\ref{ap:outth}).
When $\epsilon = 1-\gamma$, this equation reduces to 
equation (\ref{tins:out}).

\end{document}